\documentclass{icrc2009}

\usepackage{graphicx}   % for including figures
\usepackage[caption=false]{caption}    % for captions
\usepackage[font=footnotesize]{subfig} % subfig.sty for a double column floating figure using two subfigures
\usepackage{fixltx2e}
\usepackage{url}

\newcommand{\shorttitle}[1]%
{\markboth{Proceedings of the 31\MakeLowercase{$^{st}$} ICRC, {\L}\'{o}d\'{z} 2009}{#1} }
\newcommand{\etal}{\MakeLowercase{\textit{et al. }}} % "et al."

\begin{document}
\title{The VERITAS Blazar Key Science Project}

\author{\IEEEauthorblockN{Wystan Benbow\IEEEauthorrefmark{1}
			  for the VERITAS Collaboration\IEEEauthorrefmark{2}}
                            \\
\IEEEauthorblockA{\IEEEauthorrefmark{1}Harvard-Smithsonian Center for Astrophysics, Cambridge, MA, USA; wbenbow@cfa.harvard.edu}
\IEEEauthorblockA{\IEEEauthorrefmark{2}see R.A. Ong et al. (these proceedings) or http://veritas.sao.arizona.edu/conferences/authors?icrc2009}
}

% please write the preseter's name and short title (3-4 words maximum)
%    which will appear at the header of the even pages.
\shorttitle{Benbow \etal The VERITAS Blazar KSP}
\maketitle

\begin{abstract}
The VERITAS array of 12-m atmospheric-Cherenkov telescopes in southern Arizona 
is currently the world's most-sensitive
detector of astrophysical very high energy (VHE; E$>$100 GeV) $\gamma$-rays. Approximately 25 blazars 
are known to emit VHE photons, and observations of blazars are one of the VERITAS
collaboration's key science projects (KSP). More than 50 of these objects have 
been observed with the array, in most cases with the deepest-ever 
VHE exposure. This contribution summarizes the VERITAS blazar KSP,
with a focus on the upper limits measured in the blazar discovery program. 
VERITAS blazar detections are presented in detail elsewhere in these proceedings.
  \end{abstract}

\begin{IEEEkeywords}
VERITAS Gamma-ray Blazar
\end{IEEEkeywords}
 
\section{Introduction}

Active galactic nuclei (AGN) are the most numerous class of identified VHE $\gamma$-ray sources.  
These objects emit non-thermal radiation across $\sim$20 orders of magnitude in energy and
rank among the most powerful particle accelerators in the universe.
A small fraction of AGN possess strong collimated outflows (jets)
powered by accretion onto a supermassive black hole (SMBH).  VHE $\gamma$-ray emission 
can be generated in these jets,  likely in a compact region very near the SMBH event 
horizon.  Blazars, a class of AGN with jets pointed along 
the line-of-sight to the observer, are of particular interest in the VHE regime.
Approximately 25 blazars, primarily high-frequency-peaked BL\,Lacs (HBL), 
are identified as sources of VHE $\gamma$-rays,
and some are spectacularly variable on time scales  
comparable to the light crossing time of their SMBH ($\sim$2 min; \cite{HESS_2155}).  
VHE blazar studies probe the environment very near the central SMBH and
address a wide range of physical phenomena, including the accretion and jet-formation processes.
These studies also have cosmological implications, as VHE blazar data can be used
to strongly constrain primordial radiation fields (see the extragalactic background
light (EBL) constraints from, e.g., \cite{HESS_EBL1, HESS_EBL2}).

VHE blazars have double-humped spectral energy distributions (SEDs), 
with one peak at UV/X-ray energies
and another at GeV/TeV energies.  The origin of the lower-energy peak
is commonly explained as synchrotron emission from the relativistic electrons 
in the blazar jets.  The origin of the higher-energy peak is
controversial, but is widely believed to be the result of inverse-Compton scattering
of seed photons off the same relativistic electrons.  The origin of the seed photons
in these leptonic scenarios could be the synchrotron photons themselves, or photons
from an external source.  Hadronic scenarios
are also plausible explanations for the VHE emission, but generally are not favored.  

Contemporaneous multi-wavelength (MWL) observations of VHE blazars, 
can measure both SED peaks and are crucial 
for extracting science from the observations of VHE blazars.  They are used to 
constrain the size, magnetic field and Doppler factor
of the emission region, as well as to determine the origin (leptonic or hadronic) 
of the VHE emission.  In leptonic scenarios, such MWL observations are
used to measure the spectrum of high-energy electrons producing 
the emission, as well as to elucidate the nature of the seed photons.  
Additionally, an accurate measure of the 
cosmological EBL density requires accurate modeling of the blazar's 
intrinsic VHE emission that can only be performed
with contemporaneous MWL observations. 

\section{VERITAS Blazar KSP}

VERITAS began routine scientific observations with the full array in 
September 2007. The performance metrics of VERITAS include an energy resolution of $\sim$15\%, 
an angular resolution of $\sim$0.1$^{\circ}$, and a sensitivity yielding
a 5$\sigma$ detection of a 1\% Crab Nebula flux object in $<$50 hours. 
The sensitivity of VERITAS will increase by a factor of $\sim$1.2 
in Fall 2009, after the relocation of one of the telescopes,
and should not degrade over time due to an active maintenance program (e.g.
frequent mirror re-coating).  For more details about VERITAS, see \cite{Ong_ICRC}.
VERITAS observes for $\sim$750 h and $\sim$250 h 
each year during periods of astronomical 
darkness and partial moonlight, respectively. 
The moonlight observations are almost exclusively used
for blazar discovery observations, and a significant fraction of the 
dark time is used for the blazar key science project (KSP).
The VERITAS blazar KSP consists of:

\begin{itemize}
\item{A VHE blazar discovery program}
\item{A Target-of-opportunity (ToO) observation program}
\item{Multi-wavelength studies of VHE blazars}
\item{Studies of distant VHE blazars to constrain the EBL}
 \end{itemize}

A major goal of the blazar KSP is to increase
the number of identified VHE blazars such that scientific conclusions
may be drawn from many members of the VHE population, rather
than studies of a few remarkable objects.  Each year $\sim$10 blazars 
considered as potential VHE $\gamma$-ray emitters are 
selected for VERITAS discovery observations in astronomical darkness.  
This fixed-duration exposure (10 h goal for each target, 
of which 200 min is guaranteed) is supplemented by 
observations of other candidates during 
partial moonlight, resulting in more than 200 h 
per year devoted to the discovery of new VHE blazars.
The discovery program is discussed in detail later,  
and the four VHE blazars discovered by VERITAS \cite{1ES0806_letter,WCom_letter,3C66A_letter,0710_ATel}
are presented elsewhere \cite{Perkins_ICRC}.

As part of the KSP an allocation of $\sim$40 h per year 
is set aside for ToO observations of blazars.
Should this initial allocation be exhausted, additional
time can be requested from a pool of
$\sim$80 h set aside for VERITAS director's discretionary time.
This ToO component is triggered by either a VERITAS blazar discovery, 
a VHE flaring alert ($>$2 Crab) from the Whipple 10-m monitoring 
of known VHE blazars, or a flaring alert at lower energies from other observatories 
(optical, X-ray or Fermi-LAT).  All four VERITAS discoveries have 
triggered ToO observations ($\sim$20 h each) to enable a better measurement
of the spectrum and light curve. These ToO data
are particularly important in the case of a VHE discovery (e.g. RGB\,J0710+591)
during moonlight observations, such that potential systematic
effects in the data can be ruled out.  The Whipple 10-m blazar
monitoring program has triggered one ToO observation.  Here,
bright flaring of Mkn\,421 was discovered in early 2008
triggering deep ($\sim$40 h) VERITAS observations.  Six
VERITAS snapshot exposures ($<$5 h) have also been triggered,
primarily by Fermi-LAT, but also by an active optical flux monitoring program.  

The MWL component of the blazar KSP consists of both pre-planned 
and ToO campaigns on VHE blazars. For the pre-planned aspect, one 
VHE blazar is selected each year (e.g. 1ES\,2344+514
in 2007-08 \& 1ES\,1218+304 in 2008-09) to receive a $\sim$30 h 
exposure that is coordinated in advance with extensive
X-ray, optical/UV and radio observations.
For the ToO MWL program, proposals triggered by either 
a VERITAS discovery or a Whipple 10-m flaring alert 
are submitted each year to major X-ray, optical 
and radio observatories. Each blazar discovered by VERITAS,
as well as the aforementioned Whipple 10-m alert for Mkn\,421,
has triggered successful MWL observations.
Highlights of the VERITAS MWL observation
program are given elsewhere in these proceedings \cite{Grube_ICRC}.

The blazar KSP also has a program to measure the EBL via 
deep observations of distant, hard-spectrum VHE blazars.
The relatively distant HBL 1ES\,1218+304 was observed 
(see, e.g., \cite{1ES1218_paper,1ES1218_ICRC})
extensively to generate these EBL constraints.
Unfortunately there are not many  
blazars similar to 1ES\,1218+304 known in the Northern Hemisphere.  Thus the EBL program
is, in part, incorporated into the VERITAS blazar discovery program
via the inclusion of distant targets. 

\section{Discovery Program}

During the first years of VERITAS' full-scale observations,
the discovery of new VHE blazars was a primary focus of the blazar
KSP.  The targets observed were largely HBL,
but also included intermediate-frequency-peaked
(IBL) and low-frequency-peaked BL\,Lac objects (LBL), as well as 
flat spectrum radio quasars (FSRQs), in an attempt to 
increase the types of blazars known to emit VHE $\gamma$-rays.
The blazar discovery targets were drawn from 
a "target list" containing objects visible
 to the telescopes at reasonable zenith angles ($-8^{\circ} < \delta < 72^{\circ}$).
No blazar with a previously published VHE limit below the sensitivity of a
typical VERITAS exposure ($\sim$10 h) was included in this target list. 
Since VHE $\gamma$-rays from distant extragalactic sources are strongly attenuated by 
interactions on EBL photons, only a few objects having a large 
($z>0.3$) were included in the target list.  All other large $z$ blazars, or
those without a measured redshift, meeting the selection criteria described later 
were excluded. After considering visibility, redshift, 
and pre-existing VHE-exposure constraints, the VERITAS target 
list contains $\sim$50 suitable blazars.  Many of these candidates 
were already observed by VHE instruments, 
but the pre-existing VHE flux limits are generally well above (factor of $\sim$10) 
the flux sensitivity of VERITAS. 

The target list is compiled largely from nearby HBL and IBL recommended as potential
VHE emitters in three separate publications \cite{CG_2001, perlman, stecker_xbl}.
The criteria of  Costamante \& Ghisellini (\cite{CG_2001}; henceforth, CG02) 
select blazars ($z<0.45$) that are bright at both X-ray and radio wavelengths, where the X-ray emission
can be attributed to synchrotron emission from relativistic electrons in
the blazar jet.  The criteria of the other two publications \cite{perlman,stecker_xbl} are
similar, effectively requiring a bright synchrotron X-ray flux and low redshift. 
However each uses different surveys, minimum X-ray fluxes, and maximum redshifts, 
resulting in overlapping, but distinct, candidate lists. All recommended candidates (i.e. those not yet detected) 
are included in the target list. As the most recent of the three aforementioned candidate
lists is now $\sim$7 years old, many nearby HBL have since been 
identified in new surveys. The X-ray brightest HBL in the recent Sedentary \cite{sedentary}
and ROXA \cite{ROXA} surveys meet the CG02 selection criteria and are 
also included in the VERITAS target list. To 
continue the VERITAS program of constraining the EBL via the detection of distant blazars, 
three distant BL\,Lac objects ($z > 0.3$) recommended by \cite{Costamante,CG_2001}, 
are included in the target list. More than half of the
nearby ($z<0.3$) EGRET blazars are already VHE-detected,
and those not yet VHE-detected are also included in the VERITAS target list.
It is important to note that the non-HBL AGN detected 
at VHE energies are all EGRET sources, and
were seen exclusively during recent VHE flaring episodes.
In February 2009, the Fermi-LAT released its
first catalog of MeV-GeV-bright blazars.  All nearby Fermi-LAT blazars 
not-yet-detected in the VHE band are also included in the VERITAS target list.
In addition, several FSRQ recommended as potential VHE emitters
by \cite{padovani,perlman} are also included in the target list.

\section{Results of Discovery Observations}

More than 50 VHE blazar candidates were observed by VERITAS between September 2007 and May 2009.
Results from the 2 HBL and 2 LBL discovered as VHE emitters by VERITAS \cite{1ES0806_letter,WCom_letter,3C66A_letter,0710_ATel}
are reported elsewhere \cite{Perkins_ICRC}.  Excluding these 4 discoveries,
a total of 47 candidates have some exposure surviving 
data-quality selection.
The total exposure on all these candidates is $\sim$310 h live time,
yielding an average exposure of $\sim$6.5 h per candidate. The 47 candidates and
their exposures are shown in Table~\ref{blazar_targets}.
Approximately 170 h (54\%) of the total exposure is split amongst the 25 observed
HBL.  The remainder is divided amongst the 9 IBL ($\sim$80 h; 27\%),
5 LBL ($\sim$20 h, 6\%), and 8 FSRQ ($\sim$40 h, 13\%).  

\begin{figure*}[!t]
   \centerline{\subfloat[Standard Cuts]{\includegraphics[width=2.5in]{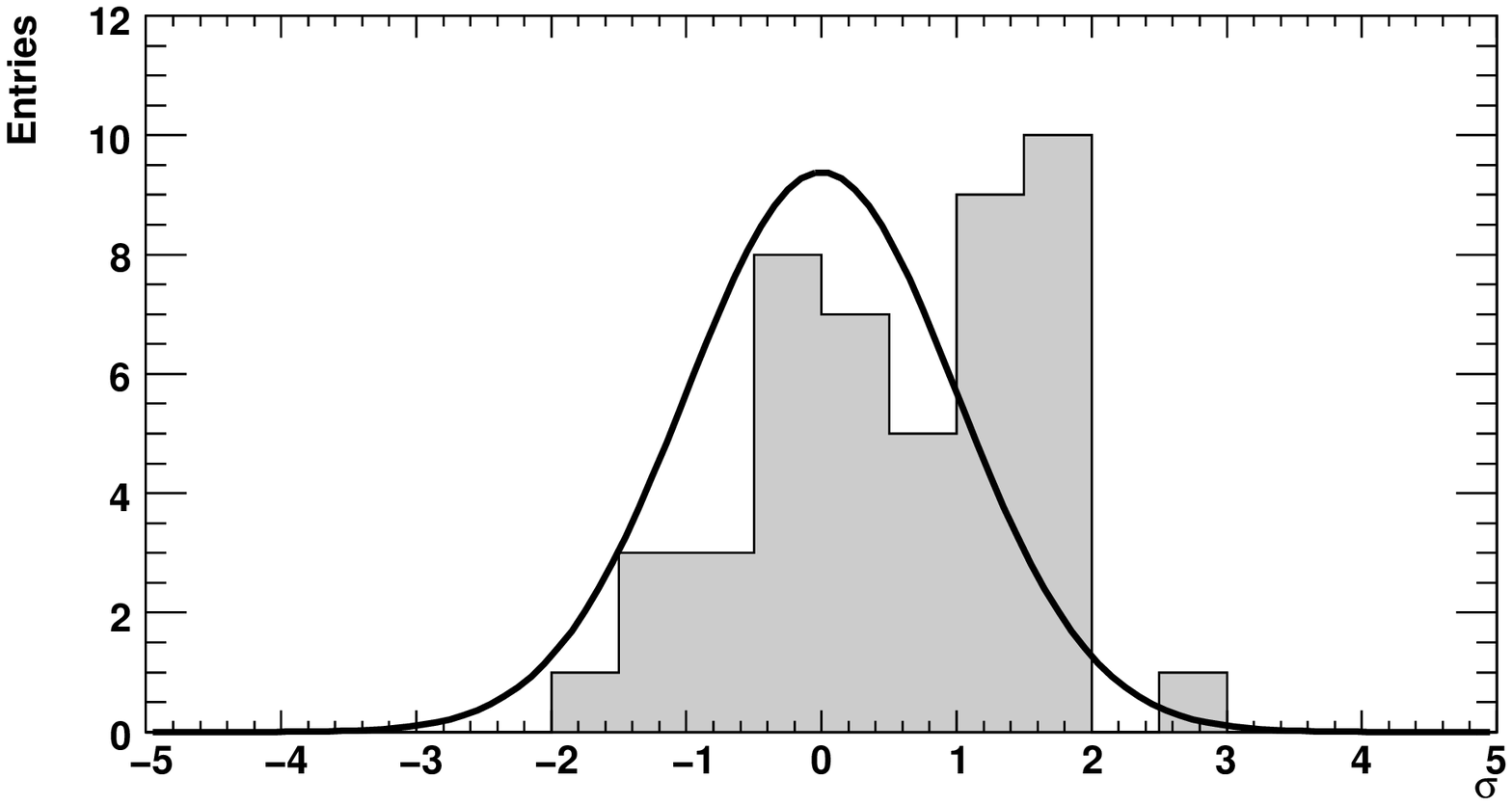} \label{sub_fig1}}
              \hfil
              \subfloat[Soft Cuts]{\includegraphics[width=2.5in]{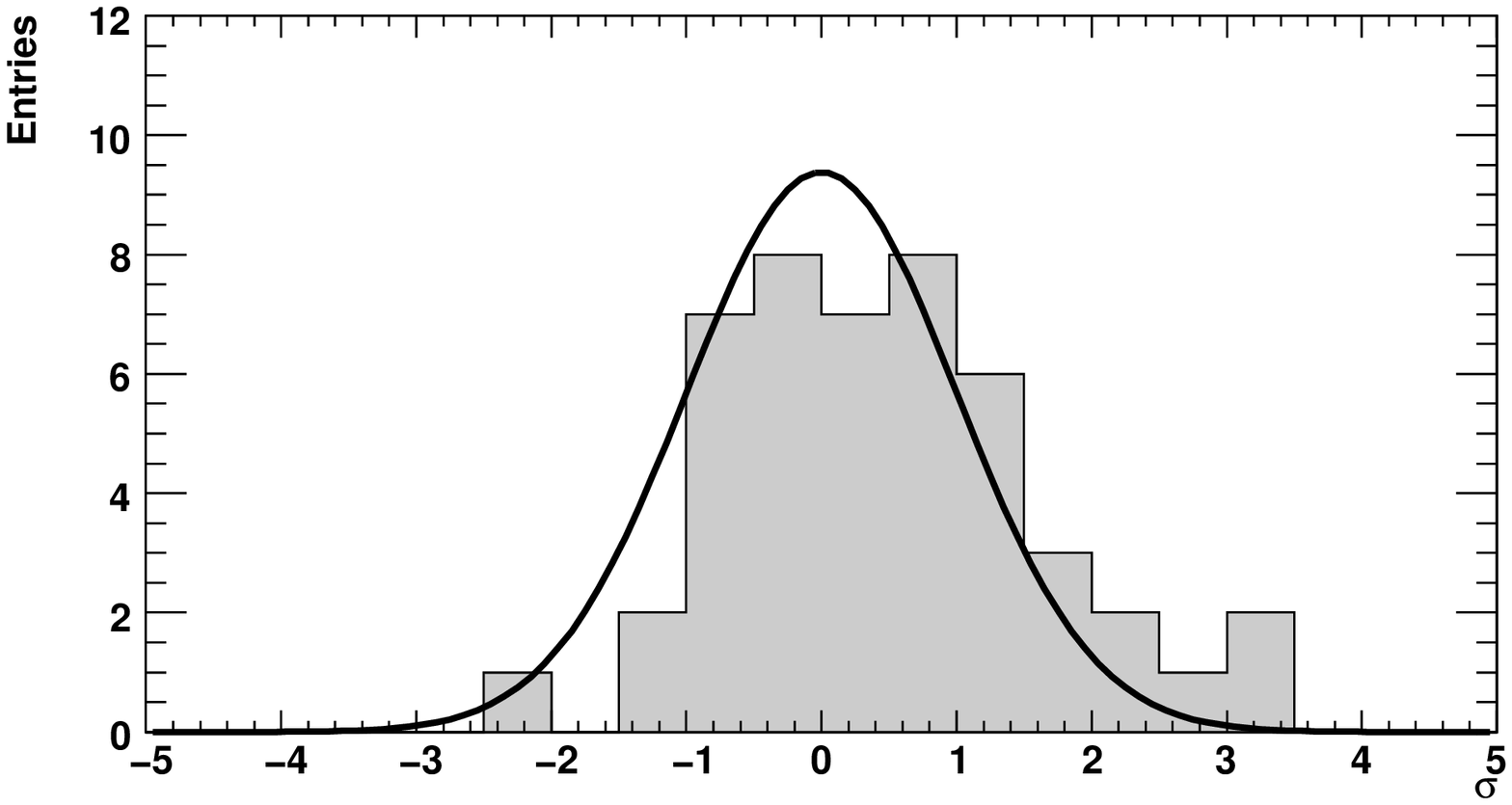} \label{sub_fig2}}
             }
   \caption{The preliminary significance measured from each of the 47 blazars. The {\it standard-cuts} (a) and {\it soft-cuts}
analysis (b) are most-sensitive to weak sources with Crab-like 
($\Gamma \approx 2.6$) and very steep (($\Gamma > 4$) spectra, respectively.  The significance observed
with each set of cuts is correlated.  The curve
shows a Gaussian distribution, with mean zero and standard deviation one, normalized
to the number of blazars.}
   \label{Sigma_blazars}
 \end{figure*}

The calibration and analysis of the VERITAS data 
is performed with the standard analysis package \cite{methods1}.
The distribution of the significance observed from each of the blazars 
is shown in Figure~\ref{Sigma_blazars} for two different, but correlated, analysis.  
The first analysis is performed with the {\it standard-cuts} 
analysis optimized using VERITAS Crab Nebula data for a weak (3\% Crab flux) source.  
The second {\it soft-cuts} analysis uses cuts that are
modified to considerably increase the sensitivity for steep-spectrum sources ($\Gamma > 4$).  
The cuts differ in the minimum image size and the size of the on-source integration region.
A small trials penalty must be accounted for due to the use of two sets of cuts.

The significance distribution for 
both analyses is skewed towards positive values
compared to what is expected from a Gaussian distribution.  Unfortunately
there are no clear outliers at positive significance which might indicate
evidence for VHE emission from a particular blazar.  The lack of
outliers is largely the result of a selection effect,
since targets which show such ''hints'' for VHE emission 
receive follow-up observations until they are
either detected or fall below a subjective significance threshold. 
Keeping this important observational bias in 
mind, a stacking analysis is performed on the entire data sample.  
Overall, there is an excess of $\sim$420 $\gamma$-rays, corresponding 
to a statistical significance of 4.8$\sigma$, observed from the directions
of the candidate blazars in the {\it standard-cuts} analysis\footnote{Stacking the {\it soft-cuts}
results yields an excess of $\sim$770 $\gamma$-rays (4.9$\sigma$).}. 
Ninety percent of the observed excess come from the HBL and IBL targets.  
This is perhaps not surprising since almost all of the known VHE blazars 
are HBL or IBL, and $\sim$80\% of the total exposure is on HBL and IBL.  

The upper limits on the VHE flux emitted by these 47 blazars 
will be presented at the conference.  
Preliminary analysis yields a typical limit of $\sim$1.5\% of the Crab Nebula flux.
The limits are almost always the most-constraining 
ever measured (see Table~\ref{blazar_targets}).  Given blazar
variability considerations, these flux limits are somewhat 
difficult to interpret.  However, these VHE limits do significantly 
constrain the low-state SED of the blazars.

  \begin{table*}[th]
  \caption{The VERITAS blazar-discovery targets.  The coordinates and redshift
are from the NED database. The BL\,Lac classification (HBL, IBL, LBL) is taken from \cite{Nieppola},
unless otherwise noted ($\dagger$ from \cite{sedentary}, LM from \cite{Laurent}, * from the BZCAT SED \cite{BZCAT}).
The VERITAS exposure is hours of good-quality live time.  The 1ES\,1215+303  exposure
is mostly from observations of 1ES\,1218+304 ($\sim$0.8$^{\circ}$ separation)
and thus has a larger average offset than the typical 0.5$^{\circ}$ wobble-mode observation.  
The notes indicate the following:  VHE recommendation from
Co \cite{Costamante}, CG \cite{CG_2001} , P \cite{perlman}, Pa \cite{padovani}, SDS \cite{stecker_xbl}; 
SS = Sedentary survey object \cite{sedentary}; R = ROXA survey object \cite{ROXA}; 
F = Fermi-LAT detection \cite{Fermi_AGN}; E = EGRET detection \cite{EGRET};  ToO = ToO trigger.
The most-constraining VHE limit, (Crab units), currently published is also shown:
HEG = HEGRA \cite{HEGRA_limit}; H = HESS \cite{HESS_limit}; M = MAGIC \cite{MAGIC_limit}; 
MIL = Milagro; W = Whipple \cite{whipple1,whipple2,whipple3}. }
  \label{blazar_targets}
  \centering
\begin{tabular}{|c|c|c|c|c|c|c|c|}
  \hline
  Source Name & RA (J2000) & Dec (J2000) & Type & Redshift & Exposure [h] & Notes & Prior VHE Limit\\
  \hline
RBS\,0042       & 00 18 27.7 & 29 47 30 & HBL$^{\dagger}$ & 0.100 & 5.5 & SS & $-$\\
1ES\,0033+595   & 00 35 52.6 & 59 50 05 & HBL & 0.086 & 23.7 & CG, P & W (11\%)\\
RGB\,J0110+418  & 01 10 04.8 & 41 49 51 & HBL & 0.096 & 2.7 & P & MIL (11\%)\\
1ES\,0120+340   & 01 23 08.6 & 34 20 49 & HBL & 0.272 & 2.1 & CG, SS & M (3.2\%)\\
QSO\,0133+476   & 01 36 58.6 & 47 51 29 & FSRQ & 0.859 & 0.8 & ToO & $-$\\ 
RGB\,J0214+517  & 02 14 17.9 & 51 44 52 & HBL & 0.049 & 3.0 & CG, P & W (17\%)\\
RBS\,0298       & 02 16 32.1 & 23 14 47 & HBL$^{\dagger}$ & 0.289 & 3.0 & SS & $-$\\
RBS\,0319 	& 02 27 16.6 & 02 02 00 & HBL$^{\dagger}$ &	0.457 & 0.3 & SS & $-$\\ 
AO\,0235+16 	& 02 38 38.9 & 16 36 59 & LBL  & 0.940 & 4.6 & F, ToO & $-$\\
RGB\,J0314+247  & 03 14 02.7 & 24 44 33 & LBL & 0.054 & 3.0 & P & MIL (18\%)\\
RBS\,0413       & 03 19 51.8 & 18 45 34 & HBL & 0.190 & 9.9 & SS & M (3.3\%)\\
1H\,0323+342 	& 03 24 41.1 & 34 10 46 & FSRQ & 0.061 & 8.3 & P & W (10\%)\\
1ES\,0414+009 	& 04 16 53.8 & 01 04 57 & HBL &	0.287 & 9.4 & CG, SS & M (5.7\%) \\  
1RXS\,J044127.8+150455 & 04 41 27.4 & 15 04 55 & HBL$^{\dagger}$ & 0.109 & 9.8  & SS & $-$\\
1ES\,0446+449   & 04 50 07.2 & 45 03 12 & IBL$^{LM}$ & 0.203 & 5.3 & SDS & $-$\\
RGB\,J0643+422 	& 06 43 26.7 & 42 14 19 & IBL$^{*}$ & 0.080 & 1.2 & & $-$\\ 
1ES\,0647+250   & 06 50 46.5 & 25 03 00 & HBL & 0.203 & 17.6 & CG & HEG 13\%\\
RGB\,J0656+426  & 06 56 10.6 & 42 37 03 & HBL & 0.059 & 9.4 & P & MIL (15\%)\\
PKS\,0829+046   & 08 31 48.9 & 04 29 39 & LBL & 0.174 & 2.4 & E & HEG (6\%)\\
Mkn\,1218 	& 08 38 10.9 & 24 53 43 & FSRQ & 0.028 & 5.9 & & W (5\%)\\ 
RGB\,J0847+115 	& 08 47 12.9 & 11 33 50 & HBL & 0.199 & 7.1 & SS & $-$\\
OJ\,287 	& 08 54 48.9 & 20 06 31 & IBL$^{*}$ & 0.306 & 6.2 & CG, E & MIL (26\%)\\ 
1ES\,0927+500   & 09 30 37.6 & 49 50 26 & HBL & 0.187 & 11.2 & R, SS & M (5.2\%)\\
1ES\,1028+511   & 10 31 18.5 & 50 53 36 & HBL & 0.360 & 11.2 & CG, R, SS & MIL (20\%)\\
RGB\,J1053+494  & 10 53 44.1 & 49 29 56 & IBL &	 0.140 & 5.6 & F & $-$\\ 	 
RBS\,0921       & 10 56 06.6 & 02 52 14 & HBL$^{\dagger}$ & 0.236 & 2.1 & SS & $-$\\
RX\,J1117.1+2014 & 11 17 06.2 & 20 14 07 & HBL$^{\dagger}$ & 0.139 & 1.8 & CG, SS, ToO & H (3.0\%)\\
RX\,J1136.5+6737  & 11 36 30.1 & 67 37 04 & HBL & 0.134 & 5.0 & CG, R, SS & MIL (54\%)\\
1ES\,1215+303   & 12 17 52.1 & 30 07 01 & IBL & 0.130? & 28.0 & CG, F  & W (22\%)\\
PKS\,1222+21    & 12 24 54.4 & 21 22 46 & FSRQ & 0.432 & 2.7 & F, ToO & $-$\\
3C\,273 	& 12 29 06.7 & 02 03 09 & FSRQ & 0.158 & 9.2 & E, F, R & $-$\\ 	 
1ES\,1255+244 	& 12 57 31.9 & 24 12 40 & HBL & 0.141 & 13.1 & SDS, SS & H (1.4\%) \\ 
RX\,J1326.2+2933 & 13 26 14.9 & 29 33 32 & IBL$^{*}$ & 0.431 & 2.4 & Co, R & $-$ \\
RGB\,J1341+399  & 13 41 04.9 & 39 59 35 & HBL & 0.163 & 0.9 & & $-$\\
RGB\,J1417+257	& 14 17 56.7 & 25 43 26 & HBL & 0.237 & 8.5 & CG, R, SS & M (2.3\%)\\ 
PKS\,1424+240 	& 14 27 00.4 & 23 48 00 & IBL & 0.160 & 7.2 & F & $-$\\ 	 
1ES\,1440+122   & 14 42 48.3 & 12 00 40 & IBL & 0.162 & 16.4 & CG, R, SDS & H (3.3\%)\\
PKS\,1510-08 	& 15 12 50.5 & -09 06 00 & FSRQ & 0.360 & 2.8 & F, ToO & $-$\\
RGB\,J1532+302  & 15 32 02.2 & 30 16 29 & HBL & 0.064 & 1.5 & P & MIL (24\%)\\
RGB\,J1610+671B & 16 10 04.1 & 67 10 26 & HBL & 0.067 & 1.2 & P & MIL (73\%)\\
1ES\,1627+402 	& 16 29 01.3 & 40 08 00 & FSRQ & 0.272 & 9.2 & Pa & W (9\%)\\ 
GB6\,J1700+6830 & 17 00 09.3 & 68 30 07 & FSRQ & 0.301 & 0.8 & F, ToO & $-$\\
PKS\,1717+177 	& 17 19 13.0 & 17 45 06 & LBL & 0.137 &	5.1 & F & $-$\\
RGB\,J1725+118 	& 17 25 04.4 & 11 52 15 & IBL & 0.018 & 9.1 & CG, P & M (4.6\%)\\
1ES\,1741+196   & 17 43 57.8 & 19 35 09 & HBL & 0.084 & 2.0 & CG & W (5.3\%) \\
RGB\,J2322+346  & 23 22 44.0 & 34 36 14 & HBL & 0.098 & 3.2 & P & MIL (25\%)\\
1ES\,2321+419   & 23 23 52.1 & 42 10 59 & LBL & 0.059 & 4.2 & SDS & HEG (3\%)\\
  \hline
  \end{tabular}
  \end{table*}

\section{Conclusion}

The first two years of the VERITAS blazar KSP
have been highly successful. Highlights include the detection
of more than a dozen VHE blazars, including 4 discoveries, with
the observations almost always having contemporaneous MWL data.  
All but a handful of the blazars on the initial 
VERITAS target list have been observed.
The excess seen in the stacked blazar analysis presented
here suggests that the direction of the VERITAS discovery program
is well justified.  Although some of the targets on the current VERITAS list
require initial or follow-up observations, a new list of candidates 
will be observed by VERITAS in future seasons.  These targets will likely be drawn from 
very hard spectrum blazars detected by Fermi-LAT, perhaps only above 1 GeV,
and will likely have a greater focus on high-risk/high-reward objects at
larger redshifts ($0.3 < z < 0.7$).  
In addition, the number of VHE blazars studied in
pre-planned MWL campains will increase as data from the
Fermi-LAT will be publically available.
In particular, future MWL observations of 1ES\,0229+200 ($z=0.139$, $\Gamma = 2.5$)
are tentatively planned given the potential EBL implications \cite{HESS_EBL2}.

\section{Acknowledgments}
{\footnotesize
This research is supported by grants from the US Department of  
Energy, the US National Science Foundation, and the Smithsonian  
Institution, by NSERC in Canada, by Science Foundation Ireland, and  
by STFC in the UK. We acknowledge the excellent work of the technical  
support staff at the FLWO and the collaborating institutions in the  
construction and operation of the instrument.
}

\end{document}